\documentclass[aps, babel,twocolumn]{revtex4}

\usepackage{amsmath,amssymb,amstext,amscd,amsthm,amsopn,
graphicx,indentfirst,mathrsfs}

\begin{document}

\title{Formation of closed timelike curves in a composite vacuum/dust 
asymptotically-flat spacetime} 
% made of vacuum and dust}
\author{Amos Ori}
 \affiliation{Department of Physics,\\
 Technion---Israel Institute of Technology, Haifa, 32000, Israel}

\begin{abstract}
We present a new asymptotically-flat time-machine model made solely of 
vacuum and dust. The spacetime evolves from a regular spacelike initial
hypersurface S and subsequently develops closed timelike curves. The initial
hypersurface S is asymptotically flat and topologically trivial. The 
chronology violation occurs in a compact manner; namely the first closed causal
curves form at the boundary of the future domain of dependence of a
compact region in S (the \textit{core}). This central core is empty, and so is
the external asymptotically flat region. The intermediate region surrounding
the core (the \textit{envelope}) is made of dust with positive energy
density. This model trivially satisfies the weak, dominant, and strong
energy conditions. Furthermore it is governed by a well-defined system of field equations 
which possesses a well-posed initial-value problem.
\end{abstract}
\maketitle

\section{Introduction}

\label{sec1}

This paper deals with the possibility of formation of closed causal curves
(CCCs) in spacetime, within the framework of General Relativity. By causal
curves we mean either timelike or null curves. The main question at the
background is the following: Is it possible that CCCs will spontaneously
evolve from rather "normal", non-pathological, initial conditions? In most
of the classic solutions of the Einstein equations (e.g. Minkowski,
Schwarzschild, Robertson-Walker) no such CCCs occur, 
although several examples of spacetimes which do admit 
closed timelike curves (CTCs) are known 
\cite{1}\cite{2}\cite{3}\cite{4}\cite{ori93}\cite{soen}\cite{ori05}\cite
{mallett}. We dub such spacetimes, which admit CTCs, as "time-machine (TM)
spacetimes". However, most of these examples suffer from pathologies or
problematic ingredients which question their relevance to physical reality.
Following is a list of three basic requirements that eliminate most of the TM
models proposed so far: (i) The spacetime should admit a regular spacelike
initial hypersurface (a partial Cauchy surface) S; (ii) asymptotic flatness;
(iii) the weak energy condition \cite{7}. Thus, Godel's rotating-dust
cosmological model \cite{1} violates conditions (i,ii), as does 
Tipler's rotating-string solution \cite{2}; the wormhole model by Moris,
Thorne and Yurtsever \cite{3} violates condition (iii). Gott's solution \cite
{4} of two infinitely long cosmic strings violates condition (ii) 
\cite{deser1}, 
and Mallett's solution \cite{mallett} violates condition (i) (see \cite{olumev}).

A few of the previous models \cite{ori93}\cite{soen}\cite{ori05} do satisfy
the above requirements (i-iii). These are all asymptotically flat,
topologically trivial models, in which the CCCs are born inside a certain
torus. These models, however, fail to satisfy some other basic requirements
discussed below. In particular, the models \cite{ori93}\cite{soen}
violate the strong energy condition \cite{7}. Here we present a new
model which better satisfies all these requirements..

In considering the physical relevance of a TM model (like any other kind of
spacetime model), one of the most important aspects is the physical
suitability of the spacetime's matter content. One certainly would like to
impose the weak energy condition, and preferably also the strong and
dominant \cite{7} energy conditions. But compliance with the energy
conditions is not enough. One would also like the spacetime's
energy-momentum tensor to coincide with some known matter field, 
governed by a well-known, well-posed field equation, for at least
two obvious reasons. First, if we cannot associate the energy-momentum
tensor with some known matter field, then we cannot tell whether this
energy-momentum distribution can be obtained in reality. Second, 
the following question concerning spacetime dynamics is at issue: Can one design
a "normal" initial configuration such that the laws of evolution will
subsequently force the spacetime to develop CCCs and to violate chronology?
A system (spacetime +matter) which does not admit a well-defined set of
evolution equations will be inadequate for addressing such a question.

Motivated primarily by the last argument, in Ref. \cite{ori05} we presented
a TM model in which the initial hypersurface is composed of three parts: An
external asymptotically flat region, the "envelope" (an intermediate
region), and a compact toroidal region at the center, to which we refer as
the "compact core". In this model CCCs evolve inside the compact core, in
a manner which is causally independent of the surrounding regions. The
central compact core, and also the external asymptotically-flat region, are
vacuum, but the envelope is made of matter. This matter satisfies all three
energy conditions mentioned above, yet the envelope's energy-momentum has
not been recognized as any known form of matter field. Nevertheless, since
the internal core is made of pure vacuum, and the formations of CCCs inside
the core is guaranteed independently of the envelope's evolution, this model
fulfills its main goal at least to some extent---it successfully
demonstrates how the laws of spacetime dynamics inevitably lead, in a
certain situation, to the violation of chronology---provided that the
initial configuration of energy-momentum at the envelope could be realized
by some real matter field. But still the question remains whether such a
matter field exists or not. One of the main goals of this paper is to
address this difficulty.

The spacetime model constructed here is similarly composed of three parts:
An internal vacuum core, an external asymptotically flat vacuum region (the
Schwarzschild geometry), and a non-empty intermediate region (the
"envelope"). Here, however, the envelope's matter will be simply\textit{\
dust} (namely, a perfect fluid with zero pressure), with non-negative energy
density. This kind of matter trivially satisfies the weak, dominant, and
strong energy conditions. It also yields a well-posed initial-value problem 
\cite{7}. Dust is probably not the most realistic or fundamental description
of matter, yet it has been proven useful in addressing various issues of
principle in General Relativity, e.g. the dust Robertson-Walker cosmology,
the Oppenheimer-Snyder model \cite{openh} of homogeneous dust collapse (the
first model to demonstrate the formation of a black hole in gravitational
collapse), and the formation of naked singularities in spherical dust
collapse \cite{crsdl}. In the last two problems, significant
progress was first made by exploring a dust model, but qualitatively similar
results were subsequently observed in models with nonvanishing pressure (see
e.g. \cite{op}). In fact, it appears that in our problem as well it will not be
difficult to generalize the present dust model to a perfect fluid with
pressure, but this is beyond the scope of this paper.

The present model also differs from that of Ref. \cite{ori05} in the type of
vacuum metric employed for the compact core. In Ref. \cite{ori05} we used a
vacuum solution locally isometric to a pp-wave \cite{exact} spacetime. Here
we use a vacuum solution locally isometric to a "{pseudo}-Schwarzschild"
geometry (namely, one obtained from the Schwarzschild geometry by a Wick
rotation), which we describe in Sec. \ref{sec3}. One of the differences
is that the {pseudo}-Schwarzschild core metric can be easily represented in
a diagonal form, which globally covers the core metric from the initial
hypersurface and up to the Cauchy Horizon (CH). We are not aware of such a
global diagonal representation of the core metric of Ref. \cite{ori05}.
Another difference is that the present core metric admits an initial
hypersurface with an especially simple extrinsic curvature, as described in
section \ref{sec3}---again, we are not aware of such a possible choice
of initial hypersurface in the core metric of Ref. \cite{ori05}. Both
factors, the diagonal core metric and the simple form of the extrinsic
curvature, greatly simplify the construction of the initial data for the TM
model with dust envelope.

There is another, more significant, difference between the two core metrics:
The locally {pseudo}-Schwarzschild metric is similar to the (four-dimensional
version of the) Misner space \cite{misner}, as its CH is entirely generated
by closed null geodesics (CNGs). In the locally-pp metric, on the other
hand, the CNGs are generically isolated. The {pseudo}-Schwarzschild core
metric is also similar to the Misner space in that two non-equivalent
analytic extensions beyond the CH exist \cite{double}. On the contrary, only
one possible extension beyond the CH appears to exist in the locally-pp core
metric of Ref. \cite{ori05}.

The Misner-like form of our present core metric has both advantages and
disadvantages. A CH generated by CNGs is claimed \cite{hawking} to be
unstable against "fragmentation" into a set of isolated null geodesics. This is
a disadvantage which might motivate us to try construct a dust envelope for
the locally-pp core of Ref. \cite{ori05} as well, but this is beyond the
scope of the present paper. At any rate we do not attempt to address
issues of stability in this paper.

But the Misner-like core used here also has advantages. In our previous
model \cite{ori05}, the question arises whether the closed null geodesic N
at the CH is adequately protected against a singularity which might form at
the future boundary of D$_{+}$(S) and approach arbitrarily close to N \cite
{future}. Originally we thought that the CNG N is causally protected against
this scenario, for the following reason: The initial hypersurface S has a
compact core S$_{0}$, such that the CNG N is located at the boundary of D$%
_{+}$(S$_{0}$). The vacuum solution in the entire region D$_{+}$(S$_{0}$) is
known explicitly, and is regular throughout. This ensures that no
singularity can form at (the boundary of) D$_{+}$(S$_{0}$) and endanger the
regularity of the neighborhood of N. However, the domain D$_{+}$(S) is
larger than D$_{+}$(S$_{0}$). Although the boundaries of these two domains
coincide at N, it may well be the case that there is
a separation between these two boundaries, and the set D$_{+}$(S)$-$D$_{+}$(S%
$_{0}$) gets arbitrarily close to N. This structure may allow the
possibility of a singularity which evolves at the future boundary of D$_{+}$
(S) (but outside D$_{+}$(S$_{0}$)) and extends arbitrarily close to N. In
that case, N would be a regular CNG (with a regular neighborhood) as viewed
from D$_{+}$(S$_{0}$), but would still lack a regular neighborhood in D$_{+}$%
(S). This might be harmful for an extended (test) observer attempting to
cross the CH (of S) through one of the points on N, in order to penetrate
into the region of CTCs. \cite{oluma}

It is not clear if this potential problem is realized in the model \cite
{ori05}. It is hard to say, because the exact solution for the
time-evolving metric is only known in the internal vacuum core and in the
external region, not in the envelope. Therefore, we do not know the full
structure of D$_{+}$(S), and, in particular, what kinds of singularities it
develops, if any. Fortunately this potential problem does not apply to the
present model. As will be shown in Sec. \ref{sec3}, due to the
Misner-like form of the CH, the boundaries of D$_{+}$(S) and D$_{+}$(S$_{0}$%
) do overlap in a set denoted H$_{1}$ below. This set, which is a portion of
the CH, includes a continuum of CNGs. A sufficiently small neighborhood of
any such CNG, restricted to D$_{+}$(S), is entirely contained in D$_{+}$(S$%
_{0}$). Since the metric throughout D$_{+}$(S$_{0}$) is known explicitly and
is perfectly smooth, no singularity which might evolve at (the boundary of) D$%
_{+}$(S) can get close to any of these CNGs.

As previously stated, the main underlying question is the
possibility of triggering the onset of chronology violation. In other words,
is it possible to design initial conditions for which the laws of dynamics
will inevitably lead to violation of chronology? But there is a build-in
logical difficulty in the formulation of this question: If indeed CTCs form,
then the portion of spacetime containing the CTCs is by definition outside
the future domain of dependence of any initial hypersurface S. In what sense
can one then state that the chronology violation has "emerged from the
initial conditions on S"? This is indeed a difficulty, but nevertheless we
propose a set of conditions which, when satisfied, provide meaning to the
statement that the violation of chronology was triggered by the initial
conditions on S. These conditions are: (i) H$_{+}$(S) contains CNGs
(therefore the Cauchy evolution of the initial data on S unambiguously leads
to some sort of chronology violation); (ii) The analytic extension of the
metric beyond H$_{+}$(S) \cite{double} (or some portion of H$_{+}$(S))
includes CTCs in the immediate neighborhood of H$_{+}$(S); 
(iii) \textit{Any} smooth extension of the metric beyond H$_{+}$(S) 
(or some portion
of H$_{+}$(S)) will include CTCs in the immediate neighborhood of H$_{+}$(S).

As a simple application of these criteria, consider the
analytically-extended geometry of a Kerr black hole. This spacetime is known
to admit CTCs deep inside the black hole. The CTCs are located beyond the
inner horizon---a null hypersurface which serves as the CH for any initial
hypersurface in the external universe. We shall not regard this spacetime as
a "time-machine model" as it fails to satisfy any of the above criteria
(i-iii). In particular, the inner horizon does not contain CNGs. On the
other hand, the model presented in this paper does satisfy all three
conditions.

When addressing the possibility of constructing a time machine, one
would primarily be interested in the situation where the construction
process takes place in a finite region of space. A simple criterion which
captures this idea is the following: We shall say that the time machine is
\textit{compactly constructed} if the initial hypersurface S includes a compact
set S$_{0}$ such that the Cauchy evolution of the initial data on S$_{0}$
leads to chronology violations; That is, the closure of D$_{+}$(S$_{0}$)
includes CCCs (specifically this means that H$_{+}$(S$_{0}$) includes CNGs).

Hawking \cite{hawking} earlier introduced a different notion of compactness called 
\textit{compact generation}. A CH is said to be compactly generated if
all its null generators, when past-propagated, enter a compact region of
spacetime and never get out of it. This criterion differs from the notion of
compact construction formulated above. The time-machine model presented in
this paper, as well as our previous models \cite{ori93}\cite{soen}\cite
{ori05}, are all compactly constructed but might not be
compactly generated.

The above discussion led to several criteria which one might apply to any
candidate model attempting to describe the process of "constructing a
time-machine" in our physical spacetime. In the next subsection we
collect these criteria and list them in a more systematic manner. Then
section \ref{sec2} outlines the structure of our model spacetime and its
various parts (central core, envelope, and external region). In Sec. \ref
{sec3} we present our core metric and discuss its main properties. Section 
\ref{sec4} outlines the initial-value setup, and the constraint equations
which must be satisfied by the initial data. Then Sections \ref{sec5} and 
\ref{sec6} describe the construction of the desired initial data (3-metric
and extrinsic curvature) on the envelope and external parts of the initial
hypersurface S, respectively. In Sec. \ref{sec7} we summarize and
discuss some of the problems and open questions remaining for future
research.

\subsection{Criteria for a physical time-machine model}

\label{sec1a}

Here we collect the various criteria which emerged in the discussion above
(plus one more criterion related to the space topology). It should be
emphasized that we do \textit{not} attempt here to postulate a strict,
formal definition of a "TM model". Rather, our goal here is to list the
various criteria which we find relevant. This list may serve as a useful
basis for discussing the physical relevance of various models which attempt
to describe "TM construction":

\begin{enumerate}

\item The spacetime should admit a spacelike initial hypersurface (a partial
Cauchy surface) S;

\item The initial data on S should be sufficiently regular; Namely, both the
spatial 3-metric and the extrinsic curvature should be C$^{(k)}$ for a
sufficiently large $k$. ($k$ should be, say, $4$ or larger in order to
guarantee a well-defined time evolution. The construction below yields C$%
^{(\infty )}$ initial data.)

\item Asymptotic flatness,

\item The spacetime's matter content should satisfy the Energy conditions.
This may be divided into two categories: (4a) The weak energy condition, and
(4b) The dominant and strong energy conditions.

\item The causal connection between S and the chronology violation: (5a) H$%
_{+} $(S) should contain CNGs; (5b) The analytic extension of the metric
beyond (some portion of) H$_{+}$(S) should include CTCs in the immediate
neighborhood of H$_{+}$(S); (5c) \textit{Any} smooth ("hole-free" \cite
{geroch}\cite{hole}) extension of the metric beyond (some portion of) H$_{+}$%
(S) should include CTCs in the immediate neighborhood of H$_{+}$(S).

\item "Causal protection" of the CNG: H$_{+}$(S) includes a CNG N admitting
a neighborhood in D$_{+}$(S) which is perfectly regular.

\item Compact construction: S should include a compact set S$_{0}$ such that H$%
_{+}$(S$_{0}$) contains CNGs. [Furthermore, the criteria 5,6 above should
apply to a portion of H$_{+}$(S) which is also contained in H$_{+}$(S$_{0}$%
).]

\item The topology of S should be trivial (R$^{3}$).

\item The energy-momentum tensor will correspond to a known matter field,
which yields a well-posed initial-value problem. This is especially crucial
for the core metric inside D$_{+}$(S$_{0}$) (which itself develops
chronology violation), but is also desired (though perhaps to a lesser
extent) for the outer parts of the time-machine model.

\end{enumerate}
We may also add the following, wider (and loosely formulated) requirement
concerning the spacetime matter content:
\begin{description}
\item[] 10. The matter field will be as elegant and/or realistic as possible. 
\end{description}
From the point of view of classical General Relativity, the most elementary and
elegant type of energy-momentum tensor is obviously the vacuum, $T_{\alpha
\beta }=0$.

The present model satisfies all these requirements. Our previous model \cite
{ori05} does not fully satisfy criterion 9, because the envelope is made
of an unrecognized matter field (though the core is vacuum). Also it is not
clear if criterion 6 is satisfied by it. Our earlier models \cite{ori93}%
\cite{soen} also fail to satisfy criterion 4b. In fact, none of the
previous models satisfy the combination of criteria 1,2,3,9.

\section{An overview of the spacetime's structure}

\label{sec2}

Our model spacetime is composed of three parts: the central active vacuum
core, the dust envelope, and the external asymptotically-flat vacuum region.
Correspondingly, the initial hypersurface S (a partial Cauchy surface) will
be composed of three parts:
\begin{enumerate}
\item The internal vacuum core, located inside a certain torus T$_{{0}}$.

\item The envelope, an intermediate region located between the torus T$_{{0}}$
and a two-sphere R$_{{s}}$ surrounding it. (R$_{{s}}$ is the two-sphere
which is later denoted $R=R_{2}$.)

\item The external vacuum region, located outside the two-sphere R$_{{s}}$.
(This external region actually corresponds to a certain spacelike
hypersurface in the Schwarzschild geometry.)
\end{enumerate}
These three parts of S will be denoted S$_{{0}}$, S$_{{1}}$, and S$_{{2}}$,
respectively.

In a similar manner we divide the evolving 4-dimensional spacetime into
three regions. To be more precise, it is the predictable portion of
spacetime, namely D$^{{+}}$(S) and its closure, which we divide and
associate with the various parts of S. The internal region M$_{{0}}$ is 
D$^{{+}}$(S$_{{0}}$); the external region M$_{{2}}$ is D$^{{+}}$(S$_{{2}}$%
), and the intermediate region M$_{{1}}$ is the intersection of D$^{{+}}$(S)
with J$^{{+}}$(S$_{{1}}$). Both M$_{{0}}$ and M$_{{2}}$ are pure vacuum
regions: M$_{{0}}$ is a compact region which includes CCCs and hence
constitutes the TM core, and M$_{{2}}$ is a portion of the Schwarzschild
geometry (the external part, which extends to spacelike infinity). The
intermediate region M$_{{1}}$ is made of dust (though it also includes
vacuum parts).

In both vacuum regions M$_{{0}}$ and M$_{{2}}$ the 4-geometry is known to us
explicitly: It is the metric (\ref{1},\ref{21}) in region M$_{{0}}$, and the
Schwarzschild geometry in region M$_{{2}}$. In the dust region M$_{{1}}$, on
the other hand, the evolving 4-geometry is not known explicitly. Instead, it
is described in terms of the corresponding initial data on S (section \ref
{sec5} below).

\section{Internal core metric}

\label{sec3}

We start our construction from the vacuum metric 
\begin{equation}
ds^{2}=-(1-2\mu /r)^{-1}dr^{2}+(1-2\mu /r)dt^{2}+r^{2}(d\theta ^{2}+\sinh
^{2}\theta d\varphi ^{2})\,,  \label{1}
\end{equation}
where $\mu $ is an arbitrary positive constant. Here $\theta $ takes all
positive values, whereas $\varphi $ admits the usual periodicity $0\leq
\varphi <2\pi $. This metric is obtained from the standard Schwarzschild
metric by a Wick rotation $\theta \rightarrow i\theta $, and we shall refer
to it as the \textit{pseudo-Schwarzschild }metric. The coordinate $t$ is
assumed here to be periodic, 
\[
0\leq t<l\,, 
\]
where $l$ is a free parameter which we take to be greater than some minimal
value $l_{\min }$ [specified in Eq. (\ref{151}) below].

The above metric has a coordinate singularity at $r=2\mu $, analogous to the
Schwarzschild's horizon (later we shall remove this singularity by
transforming to Eddington-like coordinates). At this stage, and throughout
most of this paper, we shall primarily be interested in the range $r>2\mu $.
Note that in this range $t$ is a spatial coordinate (i.e. $g_{tt}>0$) and $r$
is a time coordinate, which we take to be past-directed (namely, $r$ \textit{%
decreases} on moving from the past to the future). Thus, all hypersurfaces $%
r=const>2\mu $ are spacelike.

To overcome the coordinate singularity at $r=2\mu $ we now transform to
"Eddington-like" coordinates in the usual manner: We define 
\[
v=-(t+r^{*})\,, 
\]
where 
\[
r^{*}\equiv r+2\mu \ln (r/2\mu -1)\,. 
\]
In the $(r,v)$ coordinates the metric becomes 
\begin{equation}
ds^{2}=(1-2\mu /r)dv^{2}+2dvdr+r^{2}(d\theta ^{2}+\sinh ^{2}\theta d\varphi
^{2})\,.  \label{21}
\end{equation}
The coordinate $v$ has the same periodicity as that of $t$, namely, for
given $(r,\theta ,\varphi )$ the point $v=l$ is identified with $v=0$. \cite
{second}

\subsection{Formation of CTCs}

The closed orbits of constant $r,\theta ,\varphi $ admit the one-dimensional
line element 
\[
ds^{2}=(1-2\mu /r)dv^{2}\,. 
\]
These orbits are spacelike throughout $r>2\mu $, but they become
timelike---namely CTCs---at $r<2\mu $. Note that the metric (\ref{21})
passes $r=2\mu $ in a perfectly regular manner, as all components are
regular and 
\[
\det (g)=-r^{4}\sinh ^{2}\theta \, 
\]
is nonvanishing there.

The "hypersurface" $r=0$ is a true, timelike, curvature singularity. Our
analysis throughout this paper is restricted to the range $r>0$.

The hypersurface $r=2\mu $ is null, and its generators are the curves of
constant $\theta ,\varphi $, which are all CNGs. This hypersurface is in
fact the CH for any partial Cauchy surface $r=const>2\mu $. It also serves
as the {chronology horizon} for the metric (\ref{21}). Namely, all points at 
$0<r<2\mu $ sit on CTCs (e.g. the curves of constant $r,\theta ,\varphi $),
but none of the points at $r>2\mu $ do (because the region $r>2\mu $ is
foliated by the spacelike hypersurfaces $r=const$).

\subsection{Initial hypersurface for the internal core}

When discussing the initial-value problem for the above spacetime, we shall
consider an initial hypersurface located at $r>2\mu $. It will be convenient
to express the metric in the diagonal form (\ref{1}) (the coordinate
singularity at $r=2\mu $ will not pose any difficulty, as it takes place
away from the initial hypersurface).

For any spacelike hypersurface $r=const>2\mu $, the spatial 3-metric is 
\begin{equation}
ds^{2}=(1-2\mu /r)dt^{2}+r^{2}(d\theta ^{2}+\sinh ^{2}\theta d\varphi
^{2})\equiv h_{ab}dx^{a}dx^{b}\,.  \label{31}
\end{equation}
Hereafter the indices $a,b$ run over the three spatial coordinates. The
extrinsic curvature $K_{a}^{b}$ of such a hypersurface is

\[
K_{t}^{t}=\frac{\mu }{r^{2}}(1-2\mu /r)^{-1/2}\,,\quad K_{\theta }^{\theta
}=K_{\varphi }^{\varphi }=r^{-1}\sqrt{(1-2\mu /r)}\,, 
\]
and all other components vanish. The two distinct eigenvalues coincide at $%
r=3\mu $ in which $K_{a}^{b}=k_{0}\delta _{a}^{b}$, or 
\begin{equation}
K_{ab}=k_{0}h_{ab}\,,  \label{41}
\end{equation}
where 
\begin{equation}
k_{0}=\frac{1}{\sqrt{27}\mu }\,.  \label{42}
\end{equation}
The form (\ref{41}) greatly simplifies the constraint equations, therefore
we shall take our initial hypersurface (for the internal core) to be at 
\[
r=3\mu \equiv r_{0}\,. 
\]
We denote this hypersurface by $\Sigma $.

For later convenience we transform the periodic coordinate $t$ of the
internal vacuum core into a new coordinate $\phi $ with a standard
periodicity 
\[
0\leq \phi <2\pi \,, 
\]
namely $\phi =(2\pi /l)t$. Then the inner-core 3-metric becomes 
\begin{equation}
ds^{2}=r_{0}^{2}(d\theta ^{2}+\sinh ^{2}\theta d\varphi ^{2})+L_{0}^{2}d\phi
^{2}\,,  \label{115}
\end{equation}
where 
\[
L_{0}=(l/2\pi )\sqrt{1-2\mu /r_{0}}=\frac{l}{2\pi \sqrt{3}}\,. 
\]

Note that the 3-metric (\ref{115}) is cylindrically-symmetric, with $\theta $
serving as the "radial" coordinate and $\varphi $ as the azimuthal
coordinate. [Later we embed this 3-metric as the core of a global
asymptotically-flat hypersurface (sections \ref{sec5},\ref{sec6} below).
This global hypersurface is axially-symmetric. It should be clarified that
it is $\phi $, not $\varphi $, which becomes the global azimuthal
coordinate.]

\subsection{Truncating the internal core metric}

The hypersurface $r=3\mu $ with the spatial 3-metric (\ref{115}) is not
asymptotically-flat. In order to match it to an asymptotically-flat
exterior, we have to truncate the internal 3-metric at a certain
two-surface. Any two-surface $\theta =const>0$ on the three-surface $r=r_{0}$
is a torus (parametrized by the two periodic coordinates $\varphi ,\phi $).
We shall truncate the initial 3-metric (\ref{115}) on such a 2-surface $%
\theta =const\equiv \theta _{0}$. We denote the portion $\theta \leq \theta
_{0}$ of the three-surface $r=r_{0}$ by S$_{{0}}$.

Consider now the set D$^{{+}}$(S$_{{0}}$), namely the future domain of
dependence of S$_{{0}}$. This set has a limited extent in the time $r$
(because none of the points at $r<2\mu $ belong to this set). At $2\mu
<r<r_{0}$, D$^{{+}}$(S$_{{0}}$) will be bounded by the null geodesics of
constant $\varphi ,\phi $ which emanate at $r=r_{0}$ from $\theta =\theta
_{0}$ and propagate towards smaller $\theta $ values. [Note that geodesics
with different $\varphi ,\phi $ will have the same orbit $\theta (r)$, due
to the cylindrical symmetry..] For our time-machine construction it is
crucial that D$^{{+}}$(S$_{{0}}$) will include a portion of the chronology
horizon at $r=2\mu $. This demand will impose a minimal value for $\theta
_{0}$, as we now discuss.

The above null orbits of constant $\varphi ,\phi $ which bound D$^{{+}}$(S$_{%
{0}}$) satisfy the differential equation 
\[
\frac{d\theta }{dr}=\frac{\sqrt{-g_{rr}}}{\sqrt{g_{\theta \theta }}}=\frac{1%
}{r\sqrt{1-2\mu /r}}\,. 
\]
The general solution of this equation is 
\begin{equation}
\theta (r)=2\ln \left[ \sqrt{r/(2\mu )}+\sqrt{r/(2\mu )-1}\right] +C_{\theta
}\,,  \label{116}
\end{equation}
where $C_{\theta }$ is an integration constant. This constant is determined
from the initial value $\theta =\theta _{0}$ at $r=3\mu $, namely 
\[
C_{\theta }=\theta _{0}-\theta _{c}\,, 
\]
where 
\[
\theta _{c}=\ln \left[ 2+\sqrt{3}\right] \cong 1.317\,. 
\]

At $r=2\mu $ the term in squared brackets in Eq. (\ref{116}) vanishes. We
therefore demand $C_{\theta }>0$. Thus we shall take the cutoff value $%
\theta _{0}$ to be $>\theta _{c}$. This ensures that the portion 
\[
0\leq \theta <\theta _{0}-\theta _{c} 
\]
of the chronology horizon at $r=2\mu $ will be included in the boundary of D$%
^{{+}}$(S$_{{0}}$). In particular this portion includes an open set of CNGs.

\subsection{The relation between H$_{+}$($\Sigma $), H$_{+}$(S$_{0}$), and H$%
_{+}$(S)}

In the non-truncated core metric (\ref{21}) the CH associated with the
(complete) initial hypersurface $\Sigma $, H$_{+}$($\Sigma $), is the
hypersurface $r=2\mu $. Once the core metric is truncated (as described
above), the structure of the CH changes. The truncated part S$_{{0}}$ has
its own CH, denoted H$_{+}$(S$_{0}$). In addition the global
asymptotically-flat initial hypersurface S (which contains S$_{0}$ as its
core) has its own CH, denoted H$_{+}$(S). Here we shall briefly discuss the
relation between these various CHs.

Consider first the structure of H$_{+}$(S$_{0}$). From the discussion in the
previous subsection it follows that this hypersurface is composed of two
parts: (i) the portion $0\leq \theta \leq \theta _{0}-\theta _{c}$ of the
hypersurface $r=2\mu $, which we denote H$_{1}$, and (ii) a null
hypersurface denoted H$_{2}$, associated with the truncation of the core
metric, whose generators emanate from the truncation 2-surface $r=3\mu
,\theta =\theta _{0}$. These generators follow the orbit (\ref{116}) (for
each $\varphi $ and $\phi $). The part H$_{1}$ is a portion of H$_{+}$($%
\Sigma $). This part (unlike H$_{2}$) is entirely generated by CNGs.

The structure of H$_{+}$(S) is more complicated and still needs be explored.
It is easy to show, however, that H$_{+}$(S) contains H$_{1}$ as a subset:
Since H$_{1}\subset $ H$_{+}$(S$_{0}$) belongs to the closure of D$_{+}$(S$%
_{0}$) (and since S$_{0}\subset $S), it must also be included in the closure
of D$_{+}$(S). But since all points of H$_{1}$ sits on CNGs, none of them
belong to D$_{+}$(S). Therefore all points of H$_{1}$ must be located at the
boundary of D$_{+}$(S) (but not on S itself), namely on H$_{+}$(S).

The fact that H$_{1}$ is contained in H$_{+}$(S) guarantees the "causal
protection" discussed in the previous sections. Consider a point P located
in H$_{1}$ but away from its intersection with H$_{2}$ (i.e. at some $\theta
<\theta _{0}-\theta _{c}$). The boundaries of D$_{+}$(S) and D$_{+}$(S$_{0}$%
) overlap in the neighborhood of N. Therefore, any sufficiently-small
neighborhood of N in the closure of D$_{+}$(S) is contained in the closure
of D$_{+}$(S$_{0}$), and is hence guaranteed to be regular. This ensures
that no singularity which evolves at the boundary of D$_{+}$(S) may get
close to P. Obviously this argument also applies to the entire CNGs located
at $\theta <\theta _{0}-\theta _{c}$.

\section{The initial-value set-up}

\label{sec4}

\subsection{Basic strategy}

Our construction of the TM spacetime is formulated in terms of the
corresponding initial data on the initial hypersurface S. These initial
data, which include the three-metric $h_{ab}$ and the extrinsic curvature $K_{ab}$ 
\cite{deser2}, 
are constructed so as to satisfy the constraint equations
(discussed below). The evolution of geometry will in turn be determined by
the evolution equations. The set of equations relevant to our model is the
Einstein-dust system, namely 
\begin{equation}
G^{\alpha \beta }=8\pi T^{\alpha \beta }=8\pi \epsilon u^{\alpha }u^{\beta
}\,,  \label{010}
\end{equation}
where $\epsilon $ is a scalar field and $u^{\alpha }$ is a normalized vector
field. These quantities correspond to the dust density and four-velocity,
respectively. The Energy conditions are all satisfied if $\epsilon \geq 0$
and $u^{\alpha }$ is a timelike vector. The system (\ref{010}), along with
the initial data on S, uniquely determine the evolution of geometry (and
matter) throughout D$_{+}$(S) \cite{7}\cite{continuity}. 

As was described in the previous section, the initial hypersurface S is
composed of three parts: The inner core S$_{0}$, the envelope S$_{1}$, and
the external region S$_{2}$. In S$_{0}$ and S$_{2}$ the initial data
correspond to vacuum (i.e. $\epsilon =0$). 
This [along with the mathematical properties of Eq. 
(\ref{010})] guarantees that the evolving geometry will be vacuum throughout 
D$_{+}$(S$_{0}$) and D$_{+}$(S$_{2}$). In the rest of D$_{+}$(S), 
$\epsilon $ will in general be positive [though it may also vanish
in certain portions of J$_{+}$(S$_{1}$)].

The evolving vacuum metric is known explicitly throughout D$_{+}$(S$_{0}$),
it is given in Eq. (\ref{21}) [or Eq. (\ref{1})]. The vacuum solution in D$
_{+}$(S$_{2}$), the external part of D$_{+}$(S), is also known analytically:
It is just the Schwarzschild solution. Our construction of S thus guarantees
that the evolving spacetime in D$_{+}$(S) will be asymptotically-flat and
will admit future null infinity. In addition, the way we construct S$_{0}$
(in particular the requirement $\theta _{0}>\theta _{c}$) guarantees that
the conditions relevant to the central core region [e.g. features (5-7) in
subsection \ref{sec1a} above] are satisfied.

The "envelope" part of the evolving spacetime is the region between D$_{+}$
(S$_{0}$) and D$_{+}$(S$_{2}$). It may be expressed as 
D$_{+}$(S)$\cap $J$^{{+}}$(S$_{{1}}$). 
The evolving metric in this part is not known to us. In
particular, we do not know which kinds of singularities (if any) develop
there, and where. This limits our present ability to analyze the full causal
structure of our spacetime, e.g. whether an event horizon form, and where
exactly is the CH (outside of D$_{+}$(S$_{0}$)). It seems that a numerical
solution of the evolution equation will be required in order to fill this
gap. Nevertheless, the known analytic vacuum solutions throughout D$_{+}$(S$%
_{0}$ ) and D$_{+}$(S$_{2}$), along with the properties of S and the initial
data on it, guarantee that all the conditions (1-10) in subsection \ref
{sec1a} are satisfied by our spacetime.

\subsection{The constraint equations}

The initial data $h_{ab}$ and $K_{ab}$, to be specified on S, are subject to
four constraint equations, which correspond to four combinations of the
Einstein tensor that are completely determined by $h_{ab}$ and $K_{ab}$. Let 
$N^{\alpha }$ be a normalized timelike vector (defined on S) orthogonal to
S, and let $x^{a}$ be a set of three spacelike coordinates parametrizing S.
Then the four constrained components of the Einstein tensor are 
\[
\hat{G}_{a}\equiv G_{a\alpha }N^{\alpha }=K_{a:b}^{b}-K_{b:a}^{b} 
\]
and 
\[
\hat{G}\equiv G_{\alpha \beta }N^{\alpha }N^{\beta }=\frac{1}{2}\left[
R^{(3)}+(K_{a}^{a})^{2}-K_{ab}K^{ab}\right] \,, 
\]
where $R^{(3)}$ is the Ricci scalar associated with the 3-metric $h_{ab}$,
and a colon denotes covariant differentiation with respect to $h_{ab}$.
(Indices of $K$ are rased and lowered with the three-metric $h_{ab}$.)

The Einstein equation $G_{\alpha \beta }=8\pi T_{\alpha \beta }$ then
imposes the three \textit{Momentum equations} 
\[
K_{a:b}^{b}-K_{b:a}^{b}=8\pi T_{a\alpha }N^{\alpha }\equiv 8\pi \hat{T}_{a} 
\]
and the \textit{Energy equation} 
\[
R^{(3)}+(K_{a}^{a})^{2}-K_{ab}K^{ab}=16\pi T_{\alpha \beta }N^{\alpha
}N^{\beta }\equiv 16\pi \hat{T}\,. 
\]
In our dust model $T^{\alpha \beta }=\epsilon u^{\alpha }u^{\beta }$. To
simplify the analysis we now choose the initial dust velocity $u^{\alpha }$
to coincide with $N^{\alpha }$. Then $\hat{T}_{a}$ vanishes (because $%
N_{a}=0 $), and $\hat{T}=\epsilon $. We obtain the momentum equation 
\begin{equation}
K_{a:b}^{b}-K_{b:a}^{b}=0  \label{101a}
\end{equation}
and the Energy equation 
\begin{equation}
R^{(3)}+(K_{a}^{a})^{2}-K_{ab}K^{ab}=16\pi \epsilon \,.  \label{102a}
\end{equation}

In the internal part S$_{{0}}$ the extrinsic curvature takes the simple form
(\ref{41},\ref{42}). We shall now adopt this same form of $K_{ab}$ for the
envelope S$_{{1}}$ as well. Then in both S$_{{0}}$ and S$_{{1}}$ the
Momentum equation (\ref{101a}) is trivially satisfied, and the only
nontrivial constraint equation is the Energy equation, which now reads 
\begin{equation}
R^{(3)}+\frac{2}{9\mu ^{2}}=16\pi \epsilon \,.  \label{102b}
\end{equation}
In the internal vacuum region S$_{{0}}$ this equation reduces to 
\begin{equation}
R^{(3)}=-\frac{2}{9\mu ^{2}}\,,  \label{111}
\end{equation}
and one can easily verify that the 3-metric (\ref{31}) indeed satisfies this
relation. In the envelope region S$_{{1}}$, we only require $\epsilon $ to
be non-negative, therefore the Energy equation becomes an inequality 
\begin{equation}
R^{(3)}\geq -\frac{2}{9\mu ^{2}}\,.  \label{112}
\end{equation}

In the external vacuum region the extrinsic curvature will no longer take
the form (\ref{41}). In this region the initial data will need to satisfy
the Momentum equation (\ref{101a}) as well as the vacuum Energy equation,
namely 
\begin{equation}
R^{(3)}+(K_{a}^{a})^{2}-K_{ab}K^{ab}=0\,.  \label{102c}
\end{equation}

The initial data for the core S$_{{0}}$ are given in Eqs. (\ref{115}) and (%
\ref{41}). We shall now proceed to construct the initial data for S$_{{1}}$
and S$_{{2}}$ as well.

\section{Initial data for the envelope}

\label{sec5}

The envelope S$_{{1}}$ interpolates between the internal vacuum core
solution and the external Schwarzschild geometry. Throughout S$_{{1}}$ the
extrinsic curvature is given by Eqs. (\ref{41},\ref{42}), and the 3-metric
only needs to satisfy the dust inequality (\ref{112}).

This region will be further divided into three sub-regions:
\begin{description}
\item[The inner part:]  A region which extends between the torus $\theta =\theta
_{0}$ (the inner boundary of S$_{{1}}$) and a certain larger torus $\theta
=\theta _{3}>\theta _{0}$;

\item[The outer part: ] An inhomogeneous spherically-symmetric dust solution
near the outer boundary of the envelope. This sub-region extends between the
2-sphere $R=R_{s}$ bounding S$_{{1}}$ and a certain 2-sphere $R=R_{1}<R_{s}$
(which contains the torus $\theta =\theta _{3}$);

\item[The intermediate part:]  A homogeneous dust solution which extends
between the inner and outer parts (i.e. between the torus $\theta =\theta
_{3}$ and the 2-sphere $R=R_{1}$).
\end{description}
Our construction of the initial data for S$_{{1}}$ will start from the inner
part, proceed with the intermediate part, and conclude with the outer part.

\subsection{The inner part}

This region interpolates between the internal vacuum solution and a
homogeneous dust solution. At its inner boundary the 3-metric is 
\begin{equation}
ds^{2}=r_{0}^{2}(d\theta ^{2}+\sinh ^{2}\theta d\varphi ^{2})+L_{0}^{2}d\phi
^{2}\,,  \label{117}
\end{equation}
and on approaching the outer boundary it will become the flat 3-metric (\ref
{161}) below. A general form which covers both metrics (as well as the
entire region in between) is

\begin{equation}
ds^{2}=r_{0}^{2}[d\theta ^{2}+F(\theta )^{2}d\varphi ^{2}]+L(\theta ,\varphi
)^{2}d\phi ^{2}\,.  \label{120}
\end{equation}
In the various sub-layers of the envelope's inner part we shall make certain
choices for the functions $F(\theta ,\varphi )$ and $L(\theta ,\varphi )$,
so as to allow the transition from the metric (\ref{117}) to (\ref{161})
without violating the positivity of the dust energy density. We require both 
$F$ and $L$ to be smooth (namely $C^{(\infty )}$) and positive, which will
ensure the regularity of the 3-metric (\ref{120}).

Our starting point is the 3-metric (\ref{117}), namely

\begin{equation}
L=const=L_{0}\;,\;F=\sinh \theta \,,  \label{121}
\end{equation}
at $\theta <\theta _{0}$. Then in the range $\theta _{0}<\theta <\theta _{1}$
(for some $\theta _{1}>\theta _{1}^{\min }>\theta _{0}$, where $\theta
_{1}^{\min }$ is given below), we still take $L=L_{0}$ but allow a general
function $F(\theta )>0$, so the 3-metric is 
\begin{equation}
ds^{2}=r_{0}^{2}[d\theta ^{2}+F(\theta )^{2}d\varphi ^{2}]+L_{0}^{2}d\phi
^{2}\,.  \label{125}
\end{equation}
The Ricci scalar is then 
\[
R^{(3)}=-2F_{,\theta \theta }/(r_{0}^{2}F)=-\frac{2}{9\mu ^{2}}\frac{
F_{,\theta \theta }}{F}\,. 
\]
Thus, apart from $F(\theta )>0$, the function $F(\theta )$ only needs to
satisfy 
\[
F_{,\theta \theta }\leq F\,. 
\]
(Note that this becomes an equality at the vacuum region $\theta <\theta
_{0} $, where $F=\sinh \theta $.)

We want the function $F(\theta )$ to join smoothly on $F=\sinh \theta $ at $%
\theta \leq \theta _{0}$. We also want it, in view of Eq. (\ref{161}) below,
to join smoothly on $F=\theta $ at $\theta \geq \theta _{1}$. It is not
difficult to construct a function $F(\theta )$ satisfying all these
features---provided that $\theta _{1}$ is sufficiently large. In fact we take

\[
\theta _{1}^{\min }=2\theta _{0}+1\,. 
\]
This construction of $F(\theta )$ is described in Appendix A.

Next, in the range $\theta _{1}<\theta <\theta _{2}$ (for any $\theta
_{2}>\theta _{1}$), we take $F(\theta )=\theta $, so the 3-metric is 
\begin{equation}
ds^{2}=r_{0}^{2}[d\theta ^{2}+\theta ^{2}d\varphi ^{2}]+L_{0}^{2}d\phi
^{2}\,.  \label{131}
\end{equation}
This metric is flat, hence the dust inequality (\ref{112} ) is trivially
satisfied.

Next, in the range $\theta _{2}<\theta <\theta _{3}$ (for any $\theta
_{3}>\theta _{2}$) we take the 3-metric in the form 
\begin{equation}
ds^{2}=r_{0}^{2}[d\theta ^{2}+\theta ^{2}d\varphi ^{2}]+L(\theta ,\varphi
)^{2}d\phi ^{2}\,.  \label{141}
\end{equation}
One finds for this metric 
\[
R^{(3)}=-\frac{2}{9\mu ^{2}}\Delta L/L\,, 
\]
where $\Delta $ denotes the 2-dimensional flat-space Laplacian in polar
coordinates, namely 
\[
\Delta L\equiv L_{,\theta \theta }+L_{,\theta }/\theta +L_{,\varphi \varphi
}/\theta ^{2}\,. 
\]
The dust inequality (\ref{112}) then implies that $L(\theta ,\varphi )$ must
satisfy 
\[
\Delta L\leq L\,. 
\]
(along with $L>0$). We now take $L(\theta ,\varphi )$ in the form 
\[
L(\theta ,\varphi )=L_{0}+f(\hat{\theta})r_{0}\theta \cos \varphi \,, 
\]
where 
\[
\hat{\theta}=(\theta -\theta _{2})/\delta \theta \,, 
\]
and $\delta \theta \equiv \theta _{3}-\theta _{2}$. The function $f(\hat{
\theta})$ may be any smooth function satisfying the following:

\begin{description}

\item[(i)] it joins smoothly on $f=0$ at $\hat{\theta}\leq 0$ (corresponding to $%
L(\theta ,\varphi )=L_{0}$ at $\theta \leq \theta _{2}$),

\item[(ii)] it joins smoothly on $f=1$ at $\hat{\theta}\geq 1$ (corresponding to $%
L(\theta ,\varphi )=L_{0}+r_{0}\theta \cos \varphi $ at $\theta \geq \theta
_{3}$),

\item[(iii)] it is monotonous in between.

\end{description}
For this choice of $L(\theta ,\varphi )$ one finds 
\[
\Delta L=r_{0}\cos \varphi \left( \frac{3f^{\prime }(\hat{\theta})}{\delta
\theta }+\frac{\theta f^{\prime \prime }(\hat{\theta})}{\delta \theta ^{2}}
\right) \,, 
\]
where a prime denotes here a differentiation with respect to $\hat{\theta}$.
Using $\theta \leq \theta _{3}$ we obtain 
\[
\Delta L\leq \frac{r_{0}}{\delta \theta }\left( 3f_{\max }^{\prime }+\frac{
\theta _{3}}{\delta \theta }f_{\max }^{\prime \prime }\right) \,, 
\]
where $f_{\max }^{\prime }$ and $f_{\max }^{\prime \prime }$ denote the
maximal absolute values of $df/d\hat{\theta}$ and $d^{2}f/d\hat{\theta}^{2}$, 
respectively, throughout the range $0\leq \hat{\theta}\leq 1$. Noting that 
\begin{equation}
L(\theta ,\varphi )\geq L_{0}-r_{0}\theta _{3}\,,  \label{150}
\end{equation}
we conclude that the dust inequality $\Delta L\leq L$ is satisfied if $L_{0}$
is greater than some minimal value 
\begin{equation}
L_{0}^{\min }=r_{0}\theta _{3}+\frac{r_{0}}{\delta \theta }\left( 3f_{\max
}^{\prime }+\frac{\theta _{3}}{\delta \theta }f_{\max }^{\prime \prime
}\right) \,.  \label{152}
\end{equation}
We shall thus take $L_{0}$ to be $>L_{0}^{\min }$ as required, hence the
dust inequality is satisfied. This corresponds to 
\begin{equation}
l>l_{\min }\equiv 2\pi \sqrt{3}L_{0}^{\min }  \label{151}
\end{equation}
in terms of the parameters of the original core metric (\ref{1}). Note that
Eq. (\ref{150}) and the above expression for $L_{0}^{\min }$ also guarantees
that $L(\theta ,\varphi )$ is strictly positive throughout $\theta <\theta
_{3}$.

\subsection{The intermediate homogeneous part}

Next, in the range $\theta >\theta _{3}$ the 3-metric takes the form 
\begin{equation}
ds^{2}=r_{0}^{2}[d\theta ^{2}+\theta ^{2}d\varphi ^{2}]+(L_{0}+r_{0}\theta
\cos \varphi )^{2}d\phi ^{2}\,.  \label{161}
\end{equation}
(Recall that both $\varphi $ and $\phi $ admit a $2\pi $ periodicity.) This
is a flat metric in somewhat unusual coordinates. To bring it to a standard
form we perform the following coordinate transformation: 
\[
\rho =r_{0}\theta \cos \varphi +L_{0}\,,\;z=r_{0}\theta \sin \varphi \,, 
\]
and the metric becomes 
\begin{equation}
ds^{2}=d\rho ^{2}+\rho ^{2}d\phi ^{2}+dz^{2}\,.  \label{171}
\end{equation}
Note that 
\[
\rho \geq L_{0}-r_{0}\theta >L_{0}^{\min }-r_{0}\theta \,, 
\]
hence $\rho $ is positive on the torus $\theta =\theta _{3}$ and in its
neighborhood.

In summary, outside the torus $\theta =\theta _{3}$ the 3-metric is flat and
the dust has a constant positive density,

\[
\epsilon =\frac{1}{72\pi \mu ^{2}}\equiv \epsilon _{0}\,. 
\]
The flat 3-metric and the uniform extrinsic curvature (\ref{41}) indicate
that the initial data at $\theta >\theta _{3}$ are just those of the
(contracting) spatially-flat dust Robertson-Walker geometry.

\subsection{The outer part: inhomogeneous spherical dust geometry}

The outermost layer of the envelope S$_{{1}}$ will be constructed to be a
spherically-symmetric, inhomogeneous, dust region, which interpolates
between the constant density $\epsilon _{0}>0$ at $\theta >\theta _{3}$ and
the vanishing density at the Schwarzschild exterior. To this end we first
transform the flat metric (\ref{171}) into standard spherical coordinates $%
(R,\Theta ,\phi )$ by 
\[
z=R\cos \Theta \,,\,\,\rho =R\sin \Theta \,.
\]
The 3-metric becomes 
\begin{equation}
ds^{2}=dR^{2}+R^{2}d\Omega ^{2}\,,  \label{401}
\end{equation}
where $d\Omega ^{2}$ is the unit 2-sphere, 
\[
d\Omega ^{2}=d\Theta ^{2}+\sin ^{2}\Theta d\phi ^{2}\,.
\]
$R$ may be expressed directly in terms of the original toroidal coordinates $%
\theta ,\varphi $: 
\[
R^{2}=\rho ^{2}+z^{2}=L_{0}^{2}+(r_{0}\theta )^{2}+2L_{0}r_{0}\theta \cos
\varphi \,.
\]
This implies an inequality 
\begin{equation}
L_{0}-r_{0}\theta \leq R\leq L_{0}+r_{0}\theta \,.  \label{403}
\end{equation}

We now truncate the flat metric (\ref{401}) at a two-sphere $R=R_{1}$. We
take 
\begin{equation}
R_{1}>L_{0}+\theta _{3}r_{0}\,.  \label{407}
\end{equation}
This ensures, by virtue of the second inequality in Eq. (\ref{403}), that
the torus $\theta =\theta _{3}$ is entirely contained at $R<R_{1}$. The flat
metric (\ref{401}) thus holds throughout the region between the torus $%
\theta =\theta _{3}$ and the sphere $R=R_{1}$ surrounding it.

The 3-metric at $R>R_{1}$ is assumed to be spherically symmetric, and we
write it in the general form 
\begin{equation}
ds^{2}=g_{RR}(R)dR^{2}+R^{2}d\Omega ^{2}\,.  \label{411}
\end{equation}
It is convenient to substitute 
\begin{equation}
g_{RR}=[1-2M(R)/R+R^{2}/27\mu ^{2}{}]^{-1}\,.  \label{413}
\end{equation}
The Ricci scalar is then 
\begin{equation}
R^{(3)}=-\frac{2}{9\mu ^{2}}+\frac{4}{R^{2}}\frac{dM}{dR}\,,
\end{equation}
and Eq. (\ref{102b}) yields 
\begin{equation}
\epsilon =\frac{1}{4\pi R^{2}}\frac{dM}{dR}\,.  \label{431a}
\end{equation}
Therefore, to ensure non-negative dust density, the function $M(R)$ must be
a monotonously-increasing (or at least non-decreasing) one.

Consider next the boundary conditions on $M(R)$. In the homogeneous dust
region at $R<R_{1}$ the 3-metric is flat, $g_{RR}=1$, which corresponds to 
\begin{equation}
M(R)=R^{3}/54\mu ^{2}\,.  \label{434}
\end{equation}
The external boundary of the spherically-symmetric dust region is at the
two-sphere $R=R_{2}$ for some $R_{2}>R_{1}$, where the dust solution is to
be matched to a spherically-symmetric vacuum solution at $R>R_{2}$ (see next
section). From Eq. (\ref{431a}) this vacuum solution is characterized by 
\[
M=const\equiv m\,.
\]
Thus, $M(R)$ is required to be a monotonously-increasing function which
smoothly joins $M=R^{3}/54\mu ^{2}$ at $R\leq R_{1}$ and $M=m$ at $R\geq %
R_{2}$. For later convenience we also demand 
\begin{equation}
M(R)\leq R^{3}/54\mu ^{2}\,.  \label{433}
\end{equation}
It is straightforward to construct a function $M(R)$ satisfying all these
requirements, for any given $R_{2}>R_{1}$ and $m>R_{1}^{3}/54\mu ^{2}$. We
shall take 
\begin{equation}
R_{2}>2m\,.
\end{equation}

From Eq. (\ref{433}) it follows that 
\[
1-2M(R)/R+R^{2}/27\mu ^{2}\geq 1\,. 
\]
This guarantees that the 3-metric (\ref{411},\ref{413}) is regular
throughout $R_{1}\leq R\leq R_{2}$.

In the homogeneous dust region Eq. (\ref{434}) implies $2M/R=R^{2}/27\mu ^{2}
$. Applying this to $R=R_{1}$, using $R_{1}>2\theta _{3}r_{0}$ [obtained
from Eqs. (\ref{152},\ref{407})] and $\theta _{3}>\theta _{c}>1$, one finds
that $2M(R_{1})>R_{1}$. Namely, the dust solution includes spherical trapped
surfaces at $R=R_{1}$ and its neighborhood.

\subsubsection*{Non-spherical modification}

The above construction of the inhomogeneous dust region was
spherically-symmetric. However it is easy to generalize it to obtain
nonspherical configurations. This increases the space of solutions, and also
allows for new kinds of causal structures. Since this modification goes
beyond the main course of this paper, we shall only sketch it briefly here.

In the first stage, one chooses the function $M(R)$ such that at a certain
region $R_{a}<R<R_{b}$, for some $R_{1}<R_{a}<R_{b}<R_{2}$, it takes the
form 
\begin{equation}
M(R)=\alpha R^{3}/54\mu ^{2}\,,  \label{B1}
\end{equation}
where $0<\alpha <1$ is a fixed number. The 3-metric then becomes 
\begin{equation}
ds^{2}=[1+\frac{1-\alpha }{27\mu ^{2}}R^{2}{}]^{-1}dR^{2}+R^{2}d\Omega
^{2}\,.  \label{B2}
\end{equation}
This is a maximally-symmetric 3-metric of negative-curvature. Recalling the
uniform extrinsic curvature (\ref{41}), one realizes that the initial data
at $R_{a}<R<R_{b}$ correspond to a contracting, unbounded (i.e. "$k=-1$"),
Robertson-Walker solution. Indeed from Eq. (\ref{431a}) the dust density is
constant, $\epsilon =\alpha \epsilon _{0}$. One can easily arrange the
parameters $\alpha ,R_{a},R_{b}$ (and the mass function) such that the
region $R<R_{a}$ is free of trapped surfaces (namely $M(R)<R/2$).

Next, one picks a point P (corresponding to certain $R,\Theta ,\phi $) on S
somewhere at $R_{a}<R<R_{b}$, and re-express the 3-metric in spherical
coordinates centered on the point P (this is possible because the metric is
maximally-symmetric). We denote these new spherical coordinates $(\hat{R},%
\hat{\Theta},\hat{\phi})$. The 3-metric in these new coordinates still takes
the form (\ref{B2}), but with $R,\Theta ,\phi $ replaced by $\hat{R},\hat{%
\Theta},\hat{\phi}$, respectively. This 3-metric may be expressed by Eqs. (%
\ref{411},\ref{413},\ref{B1}), with $R$ and $\Omega $ replaced by $\hat{R}$
and $\hat{\Omega }$, respectively.

Finally one picks a two-sphere $\hat{R}=\hat{R}_{0}$ around P which is
entirely contained in $R_{a}<R<R_{b}$. At $\hat{R}<\hat{R}_{0}$ one modifies
the mass function and picks a (monotonously increasing) smooth function $M(%
\hat{R})$ at will. [Optionally one may also modify the extrinsic curvature,
namely replace Eq. (\ref{41}) by a more general spherically-symmetric form,
in a manner described in the next section.] One then obtains an
inhomogeneous dust solution at $\hat{R}<\hat{R}_{0}$.

This modified solution at $\hat{R}<\hat{R}_{0}$ is on itself spherically
symmetric, but is not concentric with the spherical shells at e.g. $%
R_{1}<R<R_{a}$. It therefore breaks the global spherical symmetry of the
external parts of S, which would otherwise extend all the way from spacelike
infinity to the trapped surfaces in the neighborhood of $R=R_{1}$.

We mentioned above that this modification may allow for new types of causal
structures. Here is one example: One can shape the inhomogeneous dust
solution at $\hat{R}<\hat{R}_{0}$ such that it will subsequently develop a
naked shell focusing singularity (similar to those constructed in e.g. Ref. 
\cite{crsdl}). It is not difficult to arrange that this singularity will be
globally naked. This spacetime will fail to be future asymptotically
predictable

\section{Initial data for the external vacuum region}

\label{sec6}

In the region $R_{2}\leq R\leq R_{3}$, for some $R_{3}>R_{2}$, we set the
3-metric

\begin{equation}
ds^{2}=(1-2m/R+R^{2}/27\mu ^{2})^{-1}dR^{2}+R^{2}d\Omega ^{2}\,,  \label{511}
\end{equation}
and the uniform extrinsic curvature (\ref{41}). The Ricci scalar is $%
R^{(3)}=-2/9\mu ^{2}$ and $\epsilon $ vanishes. Note that the term $%
1-2m/R+R^{2}/27\mu ^{2}$ was shown to be positive (if fact $>1$) at $R=R_{2}$
and it is also monotonously increasing in $R$, hence it is positive
throughout $R_{2}\leq R\leq R_{3}$.

Since the initial geometry in this range is both vacuum and spherically
symmetric, it must correspond to that of a certain spherically-symmetric
initial hypersurface in the Schwarzschild geometry. Although the
Schwarzschild spacetime is asymptotically-flat, the initial 3-metric (\ref
{511}) is obviously not. (In fact this three-metric is the same as that of a
time-symmetric hypersurface in the de Sitter spacetime.) As it turns out,
the initial data we have constructed in the range $R_{2}\leq R\leq R_{3}$
correspond to a "hyperbolic" (rather than time-symmetric) initial
hypersurface in the Schwarzschild geometry. Since we do want S to be
asymptotically-flat, essentially what we need is to deform this initial
hypersurface at $R>R_{3}$ (say), so that at large $R$ it approaches a
time-symmetric hypersurface in Schwarzschild, namely one with vanishing
extrinsic curvature and asymptotically-flat 3-metric.

Here we describe the construction of S in terms of the initial data for $h$
and $K$ (rather than through its embedding in a given spacetime). The
discussion above makes it obvious, though, that in order to make S
asymptotically flat we must relax the condition (\ref{41}) on $K_{ab}$ at $%
R>R_{3}$. (This would amount to "changing the embedding of S in
spacetime".)

In the next two subsections we shall construct $K_{ab}$ and $h_{ab}$,
respectively, in the range $R>R_{3}$. Our only presumption is that both
tensors are spherically-symmetric, with $h_{ab}$ given by Eq. (\ref{411}).
The extrinsic curvature will be obtained from the momentum equation (\ref
{101a}), and the 3-metric [namely the function $g_{RR}(R)$] will in turn be
derived from the vacuum Energy equation (\ref{102c}).

\subsection{The extrinsic curvature}

Being a spherically-symmetric tensor, we write $K_{ab}$ as 
\begin{equation}
K_{ab}=K_{0}(R)h_{ab}+\delta K(R)n_{a}n_{b}\,,  \label{521}
\end{equation}
where $n_{a}$ is the unit radial vector field. This expression must satisfy
the momentum equation 
\[
K_{a:b}^{b}-K_{b:a}^{b}=0\,. 
\]
Owing to the linearity of this equation, we may consider the contribution of
each term in Eq. (\ref{521}) separately. Since $K_{0}h_{b}^{b}=3K_{0}$, the
first term contributes $-2K_{0,a}$. The contribution of the second term is 
\begin{equation}
\left[ \delta K(R)n_{a}n^{b}\right] _{:b}-\left[ \delta
K(R)n_{b}n^{b}\right] _{:a}\,.  \label{801}
\end{equation}
Since the gradient of $\delta K$ is tangent to $n$, the contribution coming
from the derivative of $\delta K$ cancels out between the two terms in Eq. (%
\ref{801}). Also $(n_{b}n^{b})_{:a}$ vanishes due to normalization. In
addition the term $n^{b}n_{a:b}$ vanishes by the geodesic equation, because $%
n_{a}$ is the tangent vector to a congruence of geodesics (the radial rays).
The expression (\ref{801}) therefore reduces to $\delta Kn_{a}n_{:b}^{b}$,
and the momentum equation becomes 
\[
2K_{0,a}=\delta Kn_{a}n_{:b}^{b}\,. 
\]
The angular components trivially satisfy this equation. In evaluating the
radial component, a straightforward calculation yields 
\[
n_{:b}^{b}=(2/R)(g_{RR})^{-1/2}\,. 
\]
Since $n_{R}=(g_{RR})^{1/2}$, the momentum equation reduces to the simple
relation 
\begin{equation}
\delta K=R\frac{dK_{0}}{dR}\,.  \label{524}
\end{equation}

Note that the choice (\ref{41}) at $R\leq R_{3}$ corresponds to $%
K_{0}(R)=const=k_{0}$. Then $K_{0}(R)$ varies in the range $R_{3}\leq R\leq
R_{4}$, for some $R_{4}>R_{3}$. At $R\geq R_{4}$, the outermost layer of S$_{%
{2}}$, we choose $K_{0}(R)=0$, which yields $K_{ab}=0$ (this would
correspond to a time-symmetric hypersurface in Schwarzschild).

Thus, in the transition region $R_{3}\leq R\leq R_{4}$ we take $K_{0}(R)$ to
be any smooth function which smoothly joins on $K_{0}(R)=k_{0}$ at $R\leq
R_{3}$ and on $K_{0}(R)=0$ at $R\geq R_{4}$. The function $\delta K(R)$ is
then defined by Eq. (\ref{524}), hence the momentum equation is satisfied.

\subsection{The 3-metric}

The 3-metric at $R>R_{3}$ will be determined from the vacuum Energy equation 
\[
R^{(3)}+(K_{a}^{a})^{2}-K_{b}^{a}K_{a}^{b}=0\,. 
\]
With the substitutions (\ref{521},\ref{524}) this equation becomes 
\begin{equation}
R^{(3)}=-2K_{0}(3K_{0}+2R\frac{dK_{0}}{dR})\,.  \label{531}
\end{equation}

In the 3-metric (\ref{411}) we now set 
\begin{equation}
g_{RR}(R)\equiv [1-2\hat{M}(R)/R]^{-1}\,.  \label{551}
\end{equation}
The Ricci scalar is then found to be 
\[
R^{(3)}=\frac{4}{R^{2}}\frac{d\hat{M}}{dR}\,. 
\]
This, combined with Eq. (\ref{531}), yields a closed expression for $d\hat{M}
/dR$ in terms of $K_{0}(R)$. After integration one obtains 
\[
\hat{M}(R)=m-R^{3}(K_{0})^{2}/2\,. 
\]
The integration constant $m$ is determined from the boundary condition at $%
R=R_{3}$. Thus, in the range $R_{3}\leq R\leq R_{4}$ (and, in fact,
throughout $R>R_{2}$) the 3-metric is (\ref{411}) with 
\begin{equation}
g_{RR}=[1-2m/R+R^{2}K_{0}(R)^{2}]^{-1}\,.  \label{571}
\end{equation}
Note that $R_{3}>R_{2}>2m$, hence the 3-metric (\ref{411},\ref{571} ) is
regular throughout $R\geq R_{3}$ (its regularity at $R<R_{3}$ was already
established above).

Finally, at $R\geq R_{4}$ we have $K_{0}=0$, hence the 3-metric and
extrinsic curvature are 
\begin{equation}
ds^{2}=(1-2m/R)^{-1}dR^{2}+R^{2}d\Omega ^{2}\,  \label{575}
\end{equation}
and 
\begin{equation}
K_{ab}=0\,.  \label{577}
\end{equation}
Obviously this corresponds to a time-symmetric initial hypersurface in a
Schwarzschild spacetime with mass $m$.

\section{discussion}

\label{sec7}

We have constructed an asymptotically flat spacetime which evolves from a
regular partial Cauchy surface S and subsequently develops CCCs. The
formation of CCCs takes place in a compact region of space; that is, S
includes a compact set S$_{0}$ such that the CCCs form at the closure of 
D$_{+}$(S$_{0}$). This central region is empty, and so is the external part of
the spacetime. The intermediate region (the "envelope") is made of dust
with non-negative energy density. Although dust is not the most realistic
description of matter, it nevertheless provides a rather simple paradigm,
which proved in the past to be useful in addressing various issues of
principle in General Relativity---e.g. gravitational collapse, formation of
naked singularities, and cosmological models. Furthermore, the system of
dust+gravity is known to yield a well-posed initial-value problem \cite{7}.

The spacetime constructed in this way satisfies all the requirements (1-10)
listed in subsection \ref{sec1a}. In particular it is smooth,
asymptotically flat, and topologically trivial (to be precise, the initial
hypersurface S is of topology R$^{3}$). It trivially satisfies the energy
conditions (weak, strong, and dominant).

The vacuum core metric was taken here to be the {pseudo}-Schwarzschild
metric. We point out that we could also use the standard Schwarzschild
metric (with the coordinate $t$ identified on a circle), or even the
4-dimensional Misner space, for the core metric, and obtain a TM model with
similar properties. However, it is only the {pseudo}-Schwarzschild metric
which admits a homogeneous initial hypersurface (namely $r=3\mu $) with a
uniform extrinsic curvature (\ref{41}). This simplifies the construction of
the initial data for the envelope, because the momentum equation is
automatically satisfied. With the alternative core metrics previously mentioned,
the construction of initial data will be slightly more complicated.

Several problems and important questions are still left open. Perhaps the
most important one is the issue of \textit{stability}. A stability analysis
is beyond the scope of this paper, but there is comment worth noting. 
Although there are indications for classical and semiclassical
instabilities in various TM models (see e.g. \cite{kim}\cite{hawking}\cite
{krw}), the robustness and effectiveness of these instabilities are still
unclear \cite{kim}\cite{krasnikov2}\cite{xin}\cite{xingot}\cite{viser}.
Further research is required in order to assess the robustness and
effectiveness of the various instability phenomena. The model constructed
here may provide a more solid basis for a systematic stability analysis. 
A few of its features that could be important for a
genuine stability analysis include: (i) having a regular initial hypersurface, (ii)
asymptotic flatness, and (iii) admitting a well-posed system of evolution
equations. None of the previous TM models demonstrated all these properties.

Two other important open questions should be mentioned here:
\begin{enumerate}
\item It may turn out that the evolving spacetime includes a black hole, and
all CCCs are imprisoned inside the event horizon. In such a case the
formation of CCCs might still have crucial implications to various aspects of
the internal black-hole physics and geometry (e.g. singularity formation),
but nevertheless the external universe will not be influenced.
\item In our present construction, the initial data on S involve strong (though
finite) gravitational fields. Is it possible to create a TM spacetime of
this kind starting from \textit{weak-field} initial data on some earlier
initial hypersurface? Rephrasing this question:
is it possible to create such a TM spacetime by sending weak
gravitational waves (from past null infinity) and diluted dust shells (from
past timelike infinity) in the inward direction?
\end{enumerate}

As a first step towards addressing these questions one may numerically
evolve the initial data on S in both the future and past directions. Future
time evolution will tell us whether a black hole forms, and if it does,
whether it engulfs the CCCs. Past time evolution will probably indicate one
of the two possibilities: (a) the back-propagated fields disperses and
weakens---the dust expands and dilutes, and the gravitational field
spreads to past null infinity as weak gravitational waves, or (b) the fields
(back-) focus to form a white hole, a naked singularity, or
pathologies of some other kind. Such numerical simulations will thus answer
the questions (1,2) above at least with regards of the specific TM model
constructed here. Note, however, that even if this specific model is indeed
found to form a black hole in the future evolution, and/or a white hole in
the past evolution, it leaves open the possibility that a modified TM model
will be free of these undesired properties. I am not aware of any theorem or
argument which establishes a firm link between the formation of CCCs and the
subsequent formation of a black hole, or the presence of a white-hole etc.
to the past of S.

One might hope to gain insight into these two questions by exploring the
initial data on S for trapped and/or anti-trapped (i.e. "past-trapped")
surfaces. Consider first the issue of trapped surfaces and their relation to
black-hole formation in future time evolution. The external
spherically-symmetric vacuum region is free of trapped surfaces, because S$%
_{2}$ is restricted to $R>R_{2}$ and $R_{2}>2m$. But as previously mentioned, 
the dust region in the neighborhood of $R=R_{1}$ does include spherical
trapped surfaces. However, the role of trapped surfaces as indicators for
black hole formation is not so clear in our case. The theorems establishing
the connection between trapped surfaces and black hole formation assume
either global hyperbolicity, or lack of CTCs, or asymptotic predictability,
or similar properties. 
Here none of these properties can be assumed apriori, especially
because the spacetime in consideration is guaranteed to develop CCCs and a
CH. For example, on the basis of proposition 9.2.1 in Ref. \cite{7}, the
occurrence of trapped surfaces on S basically tells us that one of the
following two scenarios will be realized: (i) a black hole will form in D$%
_{+}$(S) and engulf the CCCs, or (ii) the CH will extend to future null
infinity, thus invalidating the condition of future asymptotic
predictability. \cite{fail}

The situation regarding anti-trapped surfaces seems to be different. 
If anti-trapped surfaces were found to be present on S, this would
provide firm evidence that in past evolution, the field cannot just
back-spread to infinity. It would indicate that D$_{-}$(S) fails to be past
asymptotically simple, meaning that some pathology must have taken place in the
past (prior to S): for example, a white hole, or a naked singularity. Thus, the
presence of anti-trapped surfaces on S would severely reduce the relevance
of the spacetime in consideration as a physical model describing the
construction of a TM. (Presumably, the future "spacetime engineers" will
not have white holes or naked singularities at their disposal.)

Fortunately, in the specific model constructed here it appears that no
anti-trapped surfaces exist on S. In regions S$_{0}$ and S$_{1}$, and also
in the part $R\leq R_{3}$ of S$_{2}$, the simple form (\ref{41}) of the
extrinsic curvature means that $K_{a}^{b}$ only has positive eigenvalues
(triple $k_{0}>0$), which does not allow for anti-trapped surfaces. In the
part $R\geq R_{3}$ of S$_{2}$ the geometry is Schwarzschild with $R>2m$,
hence again there are no anti-trapped surfaces.

Although the lack of anti-trapped surfaces is encouraging, recall that it is
a necessary but not a sufficient condition for the non-pathological
asymptotic structure of D$_{-}$(S). A numerical simulation of the initial
data on S towards both the past and future directions could therefore provide
valuable insight into this issue of weak-field initial data prior to S, as
well as into the problem of black hole formation in the future of S.

\section*{Acknowledgments}

I am grateful to Joseph Avron, Oded Kenneth, Dana Levanony, 
Sergey Krasnikov and James Vickers for interesting and valuable discussions.

\appendix

\section*{Appendix A}

We present here the construction of the function $F(\theta )$ of Eq. (\ref
{125}) in the range $\theta _{0}<\theta <\theta _{1}$, where $\theta _{0}$
and $\theta _{1}$ are given numbers satisfying $\theta _{0}>\theta _{c}$ and 
$\theta _{1}>\theta _{1}^{\min }\equiv 2\theta _{0}+1$. Throughout this
appendix a prime will denote differentiation with respect to $\theta $.

Recall the required properties of the function $F(\theta )$: 
(i) It is strictly positive, 
(ii) $F^{\prime \prime }\leq F$, 
(iii) It is smooth in the range $\theta _{0}<\theta <\theta _{1}$,
(iv) It joins smoothly on $F(\theta )=\sinh \theta $ at $\theta \leq \theta
_{0}$,
(v) It joins smoothly on $F(\theta )=\theta $ at $\theta \geq \theta _{1}$.

This construction is naturally divided into two stages: In the first one
(stage A below) we construct a function $\hat{F}(\theta )$ which satisfies
all the above requirements except that it is non-smooth ($\hat{F}^{\prime }$
is discontinuous) at two points, denoted $\hat{\theta}_{0}$ and $\hat{\theta}
_{1}$, which satisfy $\theta _{0}<\hat{\theta}_{0}<\hat{\theta}_{1}<\theta
_{1}$. Nevertheless $\hat{F}(\theta )$ is continuous at $\theta =\hat{\theta}
_{0}$ and $\theta =\hat{\theta}_{1}$. Furthermore, we make sure that the
"jump" in $\hat{F}^{\prime }(\theta )$ is negative at both points $\hat{%
\theta}_{0}$ and $\hat{\theta}_{1}$ (namely, the one-sided derivative in the
direction $\theta >\hat{\theta}_{0,1}$ is smaller than the corresponding one
in the direction $\theta <\hat{\theta}_{0,1}$). Then in the next stage
(referred to as stage B) the function $\hat{F}(\theta )$ is
modified---smoothened---in narrow neighborhoods of both $\theta =\hat{\theta}
_{0}$ and $\theta =\hat{\theta}_{1}$, to obtain a smooth function $F(\theta
) $ in the entire range $\theta _{0}<\theta <\theta _{1}$ which satisfies
all the above requirements (i-v).

Here we shall describe in some detail the procedure comprising stage A. The
main statement underlying stage B, namely, that a function with a "kink"
(of the correct sign) can be smoothened without violating properties (i,ii)
above, is quite obvious, but the full presentation of the detailed
smoothening procedure is lengthy. We shall therefore skip the detailed
description of stage B

\subsection*{Stage A: Constructing the rough function $\hat{F}(\theta )$}

Let us define 
\begin{equation}
\tilde{F}(\theta )\equiv a\cosh (\theta -2\theta _{0})\,\,.  \label{A21}
\end{equation}
We take the function $\hat{F}(\theta )$ to be $\hat{F}(\theta )=\tilde{F}%
(\theta )$ at $\hat{\theta}_{0}\leq \theta \leq \hat{\theta}_{1}$, along
with $\hat{F}(\theta )=\sinh \theta $ at $\theta \leq \hat{\theta}_{0}$ and $%
\hat{F}(\theta )=\theta $ at $\theta \geq \hat{\theta}_{1}$. The points $%
\hat{\theta}_{0}$ and $\hat{\theta}_{1}$ will thus be the intersection
points of $\tilde{F}(\theta )$ with $\sinh \theta $ and $\theta $,
respectively. (The smooth matching at points $\theta _{0}$ and $\theta _{1}$
is thus trivially satisfied, but as we mentioned above the challenge remains
to arrange the smooth matching at $\hat{\theta}_{0}$ and $\hat{\theta}_{1}$
---the central task in stage B.)

We take 
\[
\hat{\theta}_{1}=2\theta _{0}+1\,, 
\]
hence $\theta _{0}<\hat{\theta}_{1}<\theta _{1}$ as desired. The parameter $%
a $ will then be derived from the requirement of continuity at $\theta =\hat{
\theta}_{1}$, and subsequently continuity at $\theta =\hat{\theta}_{0}$ will
determine the value of $\hat{\theta}_{0}$. Note that Eq. (\ref{A21})
satisfies $\tilde{F}^{\prime \prime }=\tilde{F}$ (hence $\epsilon $
vanishes---and condition (ii) above holds---throughout the range where $F=%
\tilde{F}$).

Continuity at $\theta =\hat{\theta}_{1}$, namely $\tilde{F}(\hat{\theta}%
_{1})=\hat{\theta}_{1}$, yields 
\begin{equation}
a=(2\theta _{0}+1)/\cosh 1\,.  \label{A31}
\end{equation}
Note that 
\[
a>(2\theta _{c}+1)/\cosh 1\cong 2.35\,,
\]
hence in particular 
\[
a>1\,.
\]
Also note that 
\[
a<2\theta _{0}+1<e^{2\theta _{0}}\,.
\]

The parameter $\hat{\theta}_{0}$ is taken to be the point where $\tilde{F}
(\theta )$ intersects $\sinh \theta $, namely it satisfies

\[
a\cosh (\hat{\theta}_{0}-2\theta _{0})=\sinh \hat{\theta}_{0}\,. 
\]
This equation has a single real root: 
\[
\hat{\theta}_{0}=\theta _{0}+(1/2)\ln \left[ \frac{ae^{2\theta _{0}}+1}{
e^{2\theta _{0}}-a}\right] \,. 
\]
Note that $\hat{\theta}_{0}>\theta _{0}$ as desired. Later we shall also
need the inequality $\hat{\theta}_{0}<2\theta _{0}$. To see this, one
compares $\tilde{F}(\theta )$ to $\sinh \theta $ at the two points $\theta
=\theta _{0}$ and $\theta =2\theta _{0}$. In the former 
\[
\tilde{F}(\theta _{0})=a\cosh \theta _{0}>\cosh \theta _{0}\,>\sinh \theta
_{0}\,. 
\]
On the other hand, at $\theta =2\theta _{0}$ one obtains 
\[
\tilde{F}(2\theta _{0})=a=(2\theta _{0}+1)/\cosh 1\,, 
\]
which is to be compared to $\sinh (2\theta _{0})$. One finds (e.g.
numerically) that 
\[
(2\theta _{0}+1)/\cosh 1<\sinh (2\theta _{0}) 
\]
for any $\theta _{0}>\theta _{c}$ (the above inequality in fact holds for
any $\theta _{0}$ greater than $0.57$, whereas $\theta _{c}\cong 1.317$).
Since $\tilde{F}(\theta )$ is smaller than $\sinh \theta $ at $\theta
=2\theta _{0}$ but greater than $\sinh \theta $ at $\theta =\theta _{0}$,
the (single) intersection point $\hat{\theta}_{0}$ must be located in
between, namely 
\[
\theta _{0}<\hat{\theta}_{0}<2\theta _{0}\,. 
\]

Finally we compare the two one-sided values of $\hat{F}^{\prime }$ at each
of the matching points $\hat{\theta}_{0}$ and $\hat{\theta}_{1}$. Starting
at $\hat{\theta}_{0}$, the directional derivative corresponding to $\theta <%
\hat{\theta}_{0}$ is 
\[
\hat{F}^{\prime }=\cosh \hat{\theta}_{0}>0\,, 
\]
whereas the one corresponding to $\theta >\hat{\theta}_{0}$ is 
\[
\hat{F}^{\prime }=\tilde{F}^{\prime }=a\sinh (\hat{\theta}_{0}-2\theta
_{0})<0\,, 
\]
hence the jump in $\hat{F}^{\prime }$ is negative. Consider next the two
one-sided values of $\hat{F}^{\prime }$ at $\hat{\theta}_{1}$. For the
direction $\theta >\hat{\theta}_{1}$ we have $\hat{F}^{\prime }=1$, and for $%
\theta <\hat{\theta}_{1}$ we have 
\[
\hat{F}^{\prime }=\tilde{F}^{\prime }=a\sinh 1\,. 
\]
Substituting the value of $a$, Eq. (\ref{A31}), we get (at $\theta =\hat{
\theta}_{1}$) 
\[
\tilde{F}^{\prime }=(2\theta _{0}+1)\tanh 1>(2\theta _{c}+1)\tanh 1\cong
2.77>1\,. 
\]
We conclude that at both $\theta =\hat{\theta}_{0}$ and $\theta =\hat{\theta}
_{1}$ the jump in $\hat{F}^{\prime }$ is negative (namely, the directional
derivative corresponding to $\theta >\hat{\theta}_{0,1}$ is smaller than the
corresponding one corresponding to $\theta <\hat{\theta}_{0,1}$), as desired.

\subsection*{Stage B: Smoothening the rough function $\hat{F}(\theta )$}

In the next stage one constructs the function $F(\theta )$ in the range $%
\theta _{0}<\theta <\theta _{1}$ to be the same as $\hat{F}(\theta )$ except
at two narrow ranges, one in the neighborhood of $\theta =\hat{\theta}_{0}$
and one in the neighborhood of $\theta =\hat{\theta}_{1}$, in which one
replaces the non-smooth function $\hat{F}(\theta )$ by a smooth one. This
must be done without violating the two equalities $F>0$ and $F^{\prime
\prime }\leq F$. This procedure is quite straightforward though a bit
tedious, and we shall skip the details here.

It should be emphasized that this smoothening is only possible if the
"jump" in $\hat{F}^{\prime }$ is negative at both $\hat{\theta}_{0}$ and $%
\hat{\theta}_{1}$ (which was indeed shown above to be the case). For, only
in this case $\hat{F}(\theta )$ satisfies the condition (ii) above in a
distributional sense.


\begin{thebibliography}{99}
\bibitem{1}  K. Godel, Rev. Mod. Phys. \textbf{21}, 447 (1949).

\bibitem{2}  F. J. Tipler, Phys. Rev. D \textbf{9}, 2203 (1974).

\bibitem{3}  M. S. Morris, K. S. Thorne, and U. Yurtsever, 
Phys. Rev. Lett. \textbf{61}, 1446 (1988).

\bibitem{4}  J. R. Gott, Phys. Rev. Lett. \textbf{66}, 1126 (1991).

\bibitem{ori93}  A. Ori, Phys. Rev. Lett. \textbf{71}, 2517 (1993);
see also A. Ori and Y. Soen, Phys. Rev. D \textbf{49}, 3990 (1994).
%
% \bibitem{orisoen}  See also A. Ori and Y. Soen, Phys. Rev. D \textbf{49}, 3990 (1994).

\bibitem{soen}  Y. Soen and A. Ori, Phys. Rev. D \textbf{54}, 4858 (1996).

\bibitem{ori05}  A. Ori, Phys. Rev. Lett. \textbf{95}, 021101 (2005).

\bibitem{mallett}  R. L. Mallett, Found. Phys. \textbf{33}, 1307 (2003).
%
% \bibitem{6}  M. Alcubierre, Class. Quant. Grav. \textbf{11}, L73 (1994).

\bibitem{7}  S. W. Hawking and G. F. R. Ellis, Large scale structure of
space-time (Cambridge University Press, 1975).

\bibitem {deser1}  For an analysis of the three-dimensional variant of Gott's solution see 
S. Deser, R. Jackiw, and G. t'Hooft, Phys. Rev. Lett. \textbf{68}, 267 (1992). 

\bibitem{olumev}  K. D. Olum and A. Everett, 
Found. Phys. Lett. \textbf{18}, 379 (2005).

\bibitem{openh}  J. R. Oppenheimer and H. Snyder, 
Phys. Rev. \textbf{56}, 455 (1939).

\bibitem{crsdl}  D. Christodoulou, Commun. Math. Phys. \textbf{93}, 171 (1984).

\bibitem{op}  A. Ori and T. Piran, Phys. Rev. Lett. \textbf{59}, 2137 (1987).

\bibitem{exact}  H. Stephani, D. Kramer, M. MacCallum, C. Hoenselaers, and
E. Herlt, Exact solutions of Einstein's field equations, 2nd ed. (Cambridge
Univ. Press, Cambridge, 2003). 

\bibitem{misner}  C. W. Misner, in "Relativity Theory and Astrophysics I:
Relativity and Cosmology", edited by J. Ehlers (American Mathematical
Society, Providence, 1967).

\bibitem{double}  The "double extension" beyond the CH of the Misner space
is sometimes regarded as an "ambiguity" in the physical extension of the
spacetime. Here we hold the point of view according to which there is no
ambiguity, and the actual extension of the physical spacetime would include
both Misner extensions simultaneously. This possibility was discussed in
Ref. \cite{7}, and although there are some mathematical delicacies, it seems
that from the physical, general-relativistic, point of view this scenario of
two simultaneous extensions is the most acceptable one. But a fuller
discussion of this issue is beyond the scope of the present paper.

\bibitem{hawking}  S. W. Hawking, Phys. Rev. D \textbf{46}, 603 (1992).

\bibitem{future}  We are not concerned here about singularities which might
form near N at the "future side" of H$_{+}$(S), because the evolution
beyond H$_{+}$(S) is not uniquely determined.

\bibitem{oluma}  Here I follow a very similar argument 
made by Olum \cite{olum} several years ago, with regards of our previous models 
\cite{ori93}\cite{soen}. I was not aware of Olum's work until recently. 
I am grateful to Peter Weiss for bringing this work to my attention.

\bibitem{olum}  K. D. Olum, Phys. Rev. D \textbf{61} 124022 (2000).

\bibitem{geroch}  R. P. Geroch, "Prediction in General Relativity," in
Foundations of Spacetime Theories, Minnesota Studies in the Philosophy of
Science, Vol. 8, J. Earman, C. Glymour, and J. Stachel (eds), pp. 81-93.
Minneapolis: University of Minnesota Press (1977).

\bibitem{hole}  For a different point of view see S. Krasnikov, Class.
Quantum Grav. \textbf{19}, 4109 (2002). (gr-qc/0111054)

\bibitem{second}  Another analytic extension is obtained by defining $%
u=t-r^{*}$ and analytically extending the metric in the $(u,r,\theta
,\varphi )$ coordinates to $r<2\mu $ [this new metric takes exactly the form
(\ref{21}) with $v$ replaced by $u$]. Our attitude towards these twin
extensions is briefly described above \cite{double}. 

\bibitem{deser2}  We employ here the ADM formalism, 
see R. Arnowitt, S. Deser, and C. W. Misner in "Gravitation: An introduction to 
current research", edited by L. Witten (Wiley, New York 1962). 

\bibitem{continuity}  In particular the field equation (\ref{010})
guarantees that $u^{\alpha }$ and $\epsilon u^{\alpha }$ satisfy the
geodesic and the continuity equations, respectively.

\bibitem{kim}  S. W. Kim and K. S. Thorne, Phys. Rev. D \textbf{43}, 3929 (1991).

\bibitem{krw}  B. S. Kay, M. J. Radzikowski, and R. M. Wald, 
Commun. Math. Phys. \textbf{183}, 533 (1997).

\bibitem{krasnikov2}  S. V. Krasnikov, Phys. Rev. D \textbf{54}, 7322 (1996).

\bibitem{xin}  Li-Xin Li, Class. Quant. Grav. \textbf{13}, 2563 (1996).

\bibitem{xingot}  Li-Xin Li and J. R. Gott, Phys. Rev. Lett. \textbf{80}, 2980 (1998).

\bibitem{viser}  An interesting discussion may be found in M. Visser, "The
quantum physics of chronology protection", gr-qc/0204022.

\bibitem{fail}  In the globally-naked example discussed at the end of
section \ref{sec5}, the condition of future asymptotic predictability is 
\textit{guaranteed} to fail.
\end{thebibliography}
\end{document}